\PassOptionsToPackage{mathlines,mathrefs}{lineno}

\documentclass[]{aastex701}
\usepackage{amsmath}
\usepackage{xcolor}
\usepackage{soul}
\usepackage[mathlines,mathrefs]{lineno}

\DeclareMathOperator*{\argmax}{argmax}
\DeclareMathOperator*{\argmin}{argmin}

\begin{document}

\title{Deconvolved IXPE Polarization Maps of the Crab Nebula in a Maximum-a-Posteriori Framework}

\author[orcid=0000-0002-7752-9389]{Keisuke Uchiyama}
%\altaffiliation{Tokyo University of science}
\affiliation{Tokyo University of Science, 1-3 Kagurazuka, Shinjuku-ku, Tokyo 162-8601, Japan}
\affiliation{RIKEN Nishina Center, 2-1 Hirosawa, Wako, Saitama 351-0198, Japan}
\email[show]{keisuke.uchiyama@a.riken.jp}

\author[]{Mikio Morii}
\affiliation{Rikkyo University, 3-34-1 Nishi-Ikebukuro,Toshima-ku, Tokyo 171-8501, Japan}
\email[show]{morii.mikio@gmail.com}

\author[orcid=0000-0002-8801-6263]{Toru Tamagawa}
\affiliation{Tokyo University of Science, 1-3 Kagurazuka, Shinjuku-ku, Tokyo 162-8601, Japan}
\affiliation{RIKEN Nishina Center, 2-1 Hirosawa, Wako, Saitama 351-0198, Japan}
\affiliation{RIKEN Pioneering Research Institute, 2-1 Hirosawa, Wako, Saitama 351-0198, Japan}
\email[show]{tamagawa@riken.jp}

%\author[orcid=0000-0000-0000-0002,gname=Bosque, sname='Sur America']{Forrest Sur Am\'{e}rica}
%\altaffiliation{Las Campanas Observatory}
%\affiliation{Universidad de Chile, Department of Astronomy}
%\email{fakeemail2@google.com}

%\author[gname=Savannah,sname=Africa]{S. Africa}
%\affiliation{South African Astronomical Observatory}
%\affiliation{University of Cape Town, Department of Astronomy}
%\email{fakeemail3@google.com}

%\author{River Europe}
%\affiliation{University of Heidelberg}
%\email{fakeemail4@google.com}

%\author[0000-0000-0000-0003,sname=Asia,gname=Mountain]{Asia Mountain}
%\altaffiliation{Astrosat Post-Doctoral Fellow}
%\affiliation{Tata Institute of Fundamental Research, Department of Astronomy}
%\email{fakeemail5@google.com}

%\author[0000-0000-0000-0004]{Coral Australia}
%\affiliation{James Cook University, Department of Physics}
%\email{fakeemail6@google.com}

%\author[gname=IceSheet]{Penguin Antarctica}
%\affiliation{Amundsen–Scott South Pole Station}
%\email{fakeemail7@google.com}

%\collaboration{all}{The Terra Mater collaboration}

\begin{abstract}

We analyze the first IXPE observation of the Crab Nebula using a polarization-resolved deconvolution method based on maximum a posteriori (MAP) estimation.
The method reconstructs the polarization-resolved intensity map as a three-dimensional distribution in sky-position and polarization-angle spaces, using an instrumental response that depends on the incident direction and polarization angle of X-ray photons.
We apply the method to the IXPE data in the 2--8~keV band and compare the result with that obtained using the basic analysis.
Our result shows sharper spatial structure, making the jet shape more clearly identifiable than that in the basic analysis.
The reconstructed image shows a statistically significant tendency for the magnetic-field orientation to be perpendicular to the intensity-gradient orientation in the jet region.
Although these results are obtained by using a simplified response where systematic uncertainty remains, they demonstrate that our MAP-based deconvolution can improve the interpretability of IXPE polarization maps for nebular sources.
\end{abstract}


\section{Introduction}
Imaging X-ray polarimetry provides a direct probe of magnetic field geometry and emission mechanism in high-energy astrophysical sources.
The Imaging X-ray Polarimetry Explorer (IXPE) enables spatially resolved measurements of linear polarization for extended objects with an arcminute-scale field of view and energy coverage in a few keV range \citep{Weisskopf_2022_IXPE_Prelaunch}.
For bright nebulae such as the Crab, IXPE has already delivered detailed maps of polarization degree and polarization angle (i.e., electric vector position angle; EVPA), revealing their spatial variations and thereby constraining particle-acceleration mechanisms and magnetic field structures \citep[e.g.,][]{Mizuno_2023_PASJ_Crab, Wong_2024}.

The Crab Nebula is a bright synchrotron-emitting nebula powered by the Crab pulsar, which has a rotation period of 34~ms and exhibits pulsed emission.
The nebula shows a torus shape around the pulsar and two jets flowing outwards aligned with the torus axis.
For optically thin synchrotron emission, the observed polarization angle is perpendicular to the magnetic field orientation \citep{RybickiLightman1979}, so polarization-resolved morphology is important for understanding the magnetic field configuration and the dynamics of the jet flow.
The angular resolution of IXPE, characterized by a half-power diameter of approximately 25~arcsec, is comparable to the width of the jet.
Therefore, the jet structure is not fully resolved in the observed images.

IXPE consists of three X-ray telescopes, each with a mirror module assembly and a Detector Unit (DU) based on a gas pixel detector.
Incident X-rays from the sky are focused by the mirror module assemblies onto the gas pixel detectors, and the detection points at the focal plane are blurred by the point-spread function (PSF) of the mirror module assemblies.
In the gas pixel detector, each photon is absorbed and produces a photoelectron through the photoelectric effect.
For a linearly polarized photon, the probability distribution of the emission direction of the photoelectron depends on the photon polarization direction and the photoelectron is scattered by the gas and then the direction of the track is broadened.
%The recorded interaction point provides the sky position of the incident photon, while the photoelectron track provides polarization direction.
The inference of a sky polarization map is therefore affected by both the telescope PSF and the detector response to photoelectron directions.
This coupled response makes it difficult to infer the sky polarization map.

%Previous studies have investigated systematic effects caused by the PSF-induced mixing of polarization signals between neighboring sky locations in IXPE data \citep{Dinsmore_2024}.
%The study modeled how the telescope PSF mixes polarization signals from neighboring sky regions and corrected the resulting contamination in the observed polarization maps, thereby improving the interpretation of polarization structures.
%These methods, however, do not aim to reconstruct the underlying sky polarization distribution with improved spatial resolution.
%Previous studies have investigated the systematic effects of the PSF-induced mixing of polarization signals and polarization leakage on spatially resolved polarization measurements \citep{Dinsmore_2024}. While the PSF mixes photons originating from neighboring sky locations, polarization leakage is a distinct effect caused by correlations between errors in the reconstructed event positions and photoelectron directions. This effect can produce artificial polarization patterns around bright sources even when the incident emission is unpolarized. They developed a model to predict and correct these leakage patterns and derived sky-calibrated PSFs.
Previous studies have investigated instrumental effects associated with the IXPE PSF and polarization leakage \citep{Bucciantini2023}. Polarization leakage arises from correlations between errors in the reconstructed photon position and photoelectron direction, producing spurious polarization patterns even for intrinsically unpolarized emission.
\citet{Dinsmore_2024} developed a model to predict and correct this effect and derived sky-calibrated PSFs from in-orbit point-source observations.
\citet{Wong_2024} incorporated this model in their analysis of the Crab Nebula and reconstructed spatially resolved polarization using the Chandra intensity image.

Here, we adopt a different approach to reconstructing the polarization map. We formulate the reconstruction in spatial--polarization-angle count space and apply a Poisson likelihood to the observed binned counts. The reconstruction is regularized with smoothness priors under a maximum a posteriori (MAP) framework and does not use an external intensity image such as the Chandra image.
Since the Crab pulsar is a bright point source whose PSF wings contaminate nearby regions, we adopt a multi-component model to separate the pulsar and nebular emission, following \citet{Morii2024}.
We make the response matrix from simulations performed with \texttt{ixpeobssim} \citep{Baldini2022}.
Polarization leakage is not included in the available simulations and is therefore not accounted for in our reconstruction.
We apply the method to the first IXPE Crab observation and compare the result with that from a basic polarization analysis using the \texttt{ixpeobssim} \texttt{pcube} product, with the same event selection applied to both analyses.
We also compare the reconstructed polarization image with that reported by \citet{Wong_2024}.

\section{Observations and data processing} \label{sec:ixedata}

The Crab Nebula has been observed by IXPE at multiple epochs.
In this work, we analyze the first IXPE Crab observation (ObsID 01001099) to enable comparisons with previous analyses based on the same dataset.
The data are provided as a Level-2 event dataset through the IXPE archive at the High Energy Astrophysics Science Archive Research Center (HEASARC).\footnote{\url{https://heasarc.gsfc.nasa.gov/docs/ixpe/archive/}}
The on-source exposure of the dataset is 92.3~ks. We select events in the 2--8~keV band for our analysis.

\begin{figure*}[!htbp]
\centering
\includegraphics[width=12cm]{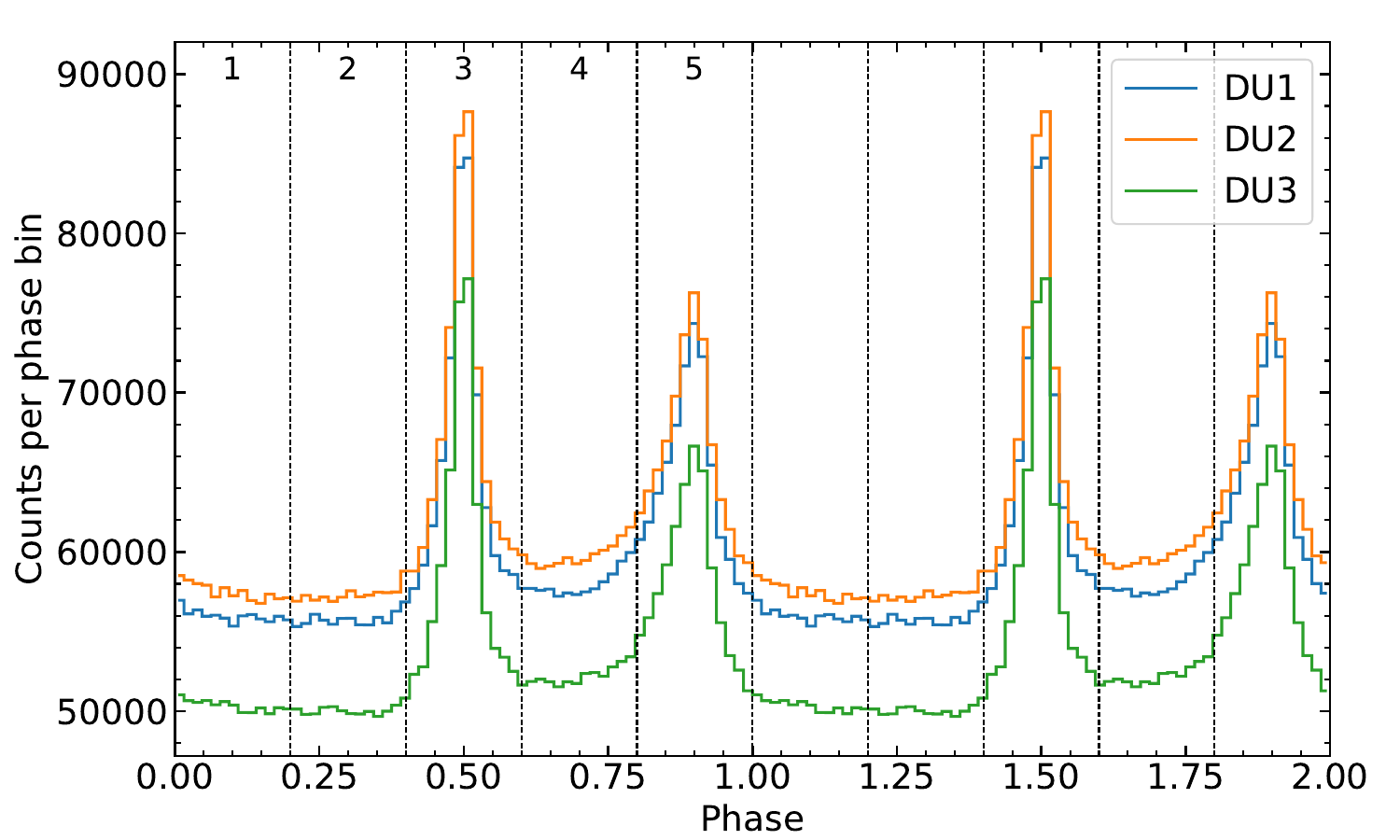}
\caption{
Folded pulse profiles (phase $0$--$2$) of the Crab pulsar in the 2--8~keV band for ObsID 01001099.
The profiles for DU1, DU2, and DU3 are shown separately (colored curves; labels in the legend).
For clarity, the pulse profile in phase $0$--$1$ is repeated over two cycles ($0$--$2$). The profiles are divided into 64 phase bins per cycle.
Dashed lines show five phase intervals of equal width (labeled 1--5).
}
\label{fig:folded_lc}
\end{figure*}

To assign pulse phases to the observed events, we apply the barycentric correction to the Level-2 event times using \texttt{barycorr} in HEASoft v6.35.1,
adopting the JPL DE440 solar-system ephemeris and the Crab pulsar coordinates recorded in the event-file header, $(\alpha,\delta)=(05^{\mathrm h}34^{\mathrm m}31.93^{\mathrm s},
+22^\circ00'52.16'')$, in the ICRS/J2000 frame.
Pulse phases are assigned to events using \texttt{xpphase} in \texttt{ixpeobssim} v31.1.1 with the Jodrell Bank Observatory (JBO) ephemeris listed in Table~\ref{tab:crab_ephem} \citep{Wong_2024}.
Figure~\ref{fig:folded_lc} shows the pulse profile obtained by folding the events. As shown by the dashed vertical lines, we divide the pulse cycle into five phase intervals of equal width and make the pulse-phase-resolved images for the intervals.
The five phase intervals are labeled 1--5 in the figure. We refer to interval~2, which has the lowest count rate, as the off-pulse phase.
For each event, the polarization angle $\phi$ is calculated from the argument of a complex number $q+{i}u$, where $q$ and $u$ are the event-by-event Stokes parameters:
$\phi = \frac{1}{2}\arg(q+{i}u)$, implemented using the \texttt{arctan2} function.

\begin{table}[!htbp]
\centering
\caption{Crab pulsar ephemeris used for pulse-phase assignment.
We adopt the JBO ephemeris reported in Table~1 of \citet{Wong_2024}.}
\label{tab:crab_ephem}
\begin{tabular}{lc}
\hline
Parameter & Value \\
\hline
PEPOCH (MJD) & 59625 \\
$\nu$ (Hz) & 29.5870753202 \\
$\dot{\nu}$ ($10^{-10}$ Hz s$^{-1}$) & $-3.67470$ \\
$\ddot{\nu}$ ($10^{-21}$ Hz s$^{-2}$) & 8.04 \\
\hline
\end{tabular}
\end{table}

\section{Method} \label{sec:mathformula}
\subsection{Mathematical formulation of the polarization-map reconstruction}
%We adopt this polarization-binned representation to formulate the inverse problem directly in count space, allowing the binned data to be modeled with a Poisson likelihood and smoothness priors to be imposed directly on the reconstructed spatial--polarization intensity distribution.

\begin{figure*}[!htbp]
\centering
\includegraphics[width=\textwidth]{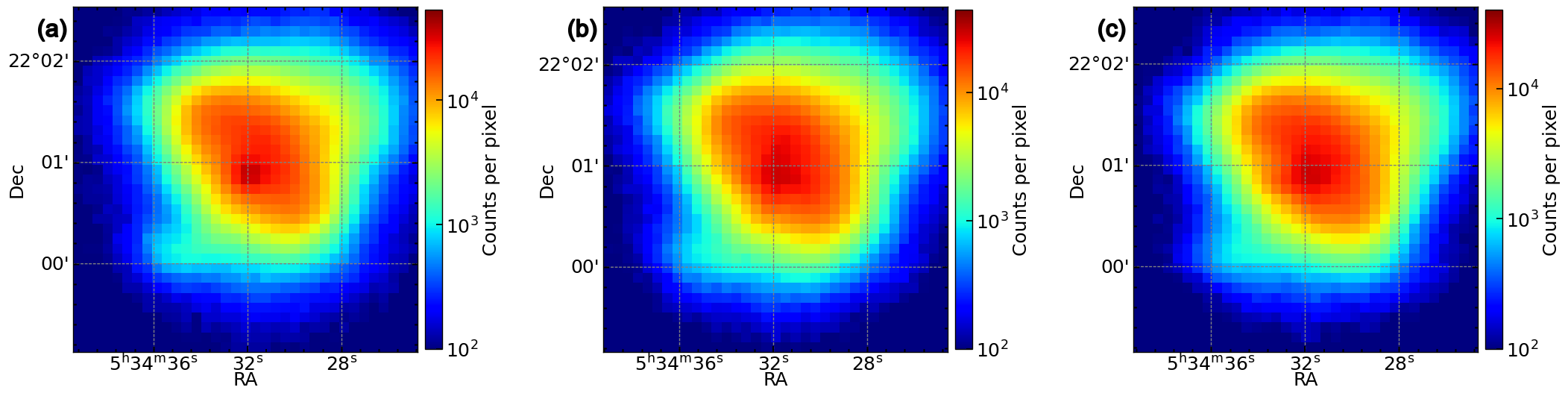}
\caption{Detector images for ObsID 01001099 in the 2--8~keV band.
The images are integrated over all pulse phases, binned with a pixel size of 5.832~arcsec, centered on the count peak in each DU, and cropped to an ROI of \(35\times35\) pixels.
Panels (a)--(c) correspond to IXPE detector units DU1--DU3,
respectively.}
\label{fig:detector_images}
\end{figure*}

We formulate the polarization-map reconstruction as an inverse problem, in which we discretize both the detector space for observed data and the sky-polarization space for the unknown intensity.
For the detector space, we bin observed events in two spatial dimensions and one
polarization-angle dimension.
In the spatial dimensions, we choose a pixel size and region of interest (ROI) to capture the Crab jet and nebular structure.
We adopt a pixel size of 5.832~arcsec and set the numbers of pixels to \(N_x=35\) and \(N_y=35\), corresponding to a
\(204.1''\times204.1''\) field of view.
For each DU, the ROI center is set to the peak position of the two-dimensional count image, which is close to the pulsar.
The adopted image centers are listed in Table~\ref{tab:center_coord},
and the resulting detector images are shown in Figure~\ref{fig:detector_images}.

\begin{table}[!htbp]
\centering
\caption{Adopted ROI center coordinates for each IXPE DU.
These centers are defined by the peak positions in the two-dimensional count images integrated over all pulse phases.
The coordinates are given in ICRS (J2000).}
\label{tab:center_coord}
\begin{tabular}{lcc}
\hline
DU & R.A. (h:m:s) & Dec. (d:m:s) \\
\hline
DU1 & 05:34:31.87 & +22:00:52.56 \\
DU2 & 05:34:31.68 & +22:00:54.72 \\
DU3 & 05:34:31.75 & +22:00:54.36 \\
\hline
\end{tabular}
\end{table}

We divide the polarization-angle interval $[-\frac{\pi}{2},\frac{\pi}{2})$ into $N_q=16$ equal-width bins.
We refer to this three-dimensional binned representation of the detector-space data as a detector cube.
It contains
$N=N_x N_y N_q=35\times35\times16$ detector voxels for each DU.
We denote the observed counts in detector voxel $t$ for DU $c$ and pulse-phase interval $p$ by $Y_t^{(c,p)}$, where $t=1,\dots,N$, \(c=1,\dots,n_c\) with \(n_c=3\), and
\(p=1,\dots,n_p\) with \(n_p=5\).

The sky-polarization space for the unknown intensity to be reconstructed is also discretized by the same spatial extent, pixel size, and polarization-angle binning as the detector space.
We refer to this discretized sky-polarization space representation as a sky cube.
The sky cube contains $N=N_xN_yN_q$ sky voxels, indexed by
$s=1,\dots,N$.
Each voxel index $s$ consists of a pair $(u,\ell)$, where
$u=(i,j)$ with $i=1,\dots,N_x$ and $j=1,\dots,N_y$ denotes the spatial pixel,
and $\ell=1,\dots,N_q$ denotes the polarization-angle bin.

For modeling a nebula containing a bright pulsar with pulse-phase dependent intensity variations, we follow \citet{Morii2024}.
We represent the sky emission as a sum of two components.
The first component is the nebular component $I_s$, which expresses the extended emission from the Crab Nebula and is assumed to be independent of pulse phase.
The second component is the point-source component
$I_{\mathrm{point},s}^{(p)}$, which expresses the contribution from the Crab pulsar and is allowed to vary with pulse phase $p$.

For each DU $c$, we define the response matrix $m^{(c)}_{t,s}$ which
gives the expected fraction of intensity at detector voxel $t$ from a
unit intensity at sky voxel $s$.
The $s$-th column of $m^{(c)}_{t,s}$ expresses the detector-space count distribution produced by a point source at sky voxel $s$.
With this response, the expected counts in detector voxel $t$ of DU $c$ and
phase interval $p$ are written as
\begin{equation}
  E_t^{(c,p)}=\Bigg[\sum_{s=1}^{N} m^{(c)}_{t,s}\,
      \Big(I_{s} + I^{(p)}_{\mathrm{point},s}\Big)\Bigg]\,d_p,
  \label{eq:mean_counts}
\end{equation}
where $d_p$ is the width of phase interval $p$ in units of pulse cycles.
In this work, we adopt equal-width phase bins, so that $d_p=1/n_p$.
The calculation of $m^{(c)}_{t,s}$ is described in Section~\ref{subsec:response}.

We model the pulsar as a point source at a fixed position $(i_{\mathrm{point}},j_{\mathrm{point}})$ and estimate only its phase- and polarization-dependent flux.
Let $f^{(q,p)}\ge0$ denote the point-source flux for polarization-angle index $q=1,\dots,N_q$ and phase index $p$.
This component includes both the phase-independent and pulsed emission from the pulsar. %(i.e., the total flux of the pulsar, including both phase-independent and pulsed components).
We write the pulsar contribution on the sky as
\begin{equation}
  I^{(p)}_{\mathrm{point},s} = \sum_{q=1}^{N_q} f^{(q,p)}\, I^{(q)}_{\mathrm{unit},s},
  \label{eq:psr_sky}
\end{equation}
where $I^{(q)}_{\mathrm{unit},s}$ is a source image with unit intensity located only at sky position $(i_{\mathrm{point}},j_{\mathrm{point}})$ and confined to the $q$-th polarization-angle bin:
\begin{equation}
  I^{(q)}_{\mathrm{unit},((i,j),\ell)} =
  \begin{cases}
    1, & ((i,j),\ell)=((i_\mathrm{point},j_\mathrm{point}),q),\\
    0, & \text{otherwise}.
  \end{cases}
\end{equation}

We assume that the observed counts $Y=\{Y^{(c,p)}_{t}:t=1,\cdots,N; c=1,\cdots,n_c; p=1,\cdots,n_p\}$ follow Poisson distributions with means $E=\{E^{(c,p)}_{t}\}$.
The likelihood is then given by
\begin{equation}
\begin{aligned}
  p\!\left(Y \mid I,f \right)
  &= \prod_{c=1}^{n_c}\prod_{p=1}^{n_p}\prod_{t=1}^{N}
     \mathrm{Poisson}\!\left(Y^{(c,p)}_{t};\, E^{(c,p)}_{t}\right) ,
\end{aligned}
\label{eq:likelihood}
\end{equation}
where $I=\{I_s\}$ and $f=\{f^{(q,p)}\}$.
Taking the logarithm and dropping terms independent of $(I,f)$ yields
\begin{equation}
\begin{aligned}
    \log p\!\left(Y \mid I,f \right)
    &= \sum_{c=1}^{n_c}\sum_{p=1}^{n_p}\sum_{t=1}^{N}
       \left( Y^{(c,p)}_{t}\,\log E^{(c,p)}_{t} - E^{(c,p)}_{t} \right).
    \end{aligned}
    \label{eq:likelihood_objective}
\end{equation}
%We adopt a maximum-a-posteriori (MAP) approach, in which we estimate $(I,f)$ by maximizing the posterior density
We adopt a maximum a posteriori (MAP) approach.
The estimated sky intensity and point-source flux, denoted by $(I^\ast,f^\ast)$, are obtained by maximizing the posterior probability using Bayes' theorem:
\begin{equation}
\begin{aligned}[b]
  (I^\ast,f^\ast)&=\argmax_{I\ge0,\,f\ge0}\; p(I,f\mid Y)\\
  &=\argmax_{I\ge0,\,f\ge0}\; p(Y\mid I,f)\;p(I,f)\\
  &=\argmax_{I\ge0,\,f\ge0} \left[\log p(Y\mid I,f)+\log p(I,f)\right],
  \end{aligned}
  \label{eq:map}
\end{equation}
where $p(I,f)$ is a prior probability.

%We model the expected counts in detector voxel $t$ for DU $c$ and phase bin $p$ as

%The instrument response describes the fraction of the intensity in each sky voxel that contributes to each detector voxel.
%For each DU $c$, we define a response matrix $m^{(c)}_{t,s}$ such that the distribution of the expected photon counts at detector voxel $t$ originating from a unit intensity of photon from sky voxel $s$.
%The calculation of $m^{(c)}_{t,s}$ is described in Section~\ref{subsec:response}.

%For modeling a nebula containing a bright pulsar with pulse-phase dependent intensity change, we follow \citet{Morii2024}.
%We divide the sky intensity into two component.
%One of it is the nebula component $I_s$ which is **広がった天体であるCrab nebulaからの寄与を意味し、pulse phase interval $p$によって変動しない成分である**
%The other is the pulsed component $I^{(p)}_{\mathrm{point},s}$ which is **点源であるCrab pulsarからの寄与を意味し、pulse interval $p$によって変動する成分である**.
%**すなわち推定するsky intensityは $I'^{(p)}_s=I_s+I^{(p)}_{\mathrm{point},s}$となる**

\subsection{Regularization }
We use three prior factors:
(i) polarization-angle smoothness of the nebular component $p_{\mathrm{smooth,pol}}$,
(ii) spatial smoothness of the nebular component $p_{\mathrm{smooth,spatial}}$, and
(iii) a constraint on the point-source flux $p_{\mathrm{flux}}$.
These factors are written as
\begin{equation}
p(I,f) \propto p_{\mathrm{smooth,pol}}(I)\, p_{\mathrm{smooth,spatial}}(I)\, p_{\mathrm{flux}}(f).
\label{eq:prior_factorization}
\end{equation}
The polarization-angle smoothness prior is
\begin{equation}
p_{\mathrm{smooth,pol}}(I)
= Z_{1} \exp\!\left[-\mu_1 V_1(I)\right],
\label{eq:prior_pol}
\end{equation}
where
\begin{equation}
 V_1(I) = \sum_{i=1}^{N_x}\sum_{j=1}^{N_y}\left[\sum_{\ell=1}^{N_q-1}(I_{(i,j),\ell} - I_{(i,j),\ell+1})^2+(I_{(i,j),1}-I_{(i,j),N_q})^2\right],
 \label{eq:V1_objective}
\end{equation}
$\mu_1$ is a hyperparameter controlling the smoothness in the polarization-angle dimension, and $Z_1$ is a normalization constant.
The last term accounts for the periodic boundary condition in the polarization-angle coordinate.
The spatial smoothness prior is
\begin{equation}
p_{\mathrm{smooth,spatial}}(I)
= Z_{2} \exp\!\left[-\mu_2 V_2(I)\right],
\label{eq:prior_spatial}
\end{equation}
where
%$V_2$は空間的に隣接したpixelのintensityの差を小さくする制約を表し、
\begin{equation}
\begin{aligned}[b]
 V_2(I)
 &= \sum_{\ell=1}^{N_q} \Biggl\{
 \sum_{i=1}^{N_x-1} \sum_{j=1}^{N_y-1}
 \left[
 \left(I_{(i,j),\ell}-I_{(i+1,j),\ell}\right)^2
 +\left(I_{(i,j),\ell}-I_{(i,j+1),\ell}\right)^2
 \right] \\
 &\quad
 +\sum_{i=1}^{N_x-1}
 \left(I_{(i,N_y),\ell}-I_{(i+1,N_y),\ell}\right)^2 \\
 &\quad
 +\sum_{j=1}^{N_y-1}
 \left(I_{(N_x,j),\ell}-I_{(N_x,j+1),\ell}\right)^2
 \Biggr\},
\end{aligned}
\label{eq:V2_objective}
\end{equation}
$\mu_2$ is a hyperparameter controlling the smoothness in the spatial dimensions, and $Z_2$ is a normalization constant.
The point-source flux prior is
\begin{equation}
p_{\mathrm{flux}}(f) = Z_3 \exp\!\left[-\mu_3 V_3(f)\right],
 \label{eq:prior_flux}
\end{equation}
where
\begin{equation}
\begin{aligned}
    V_3(f)=\sum_{p=1}^{n_p}\left( \sum_{q=1}^{N_q} f^{(q,p)}-f_{\mathrm{ref}}^{(p)} \right)^2,
    \end{aligned}
    \label{eq:V3_objective}
\end{equation}
$\mu_3$ is a hyperparameter controlling the strength of the flux constraint, and $Z_3$ is a normalization constant.
This prior encourages the point-source flux summed over the polarization-angle bins to be close to the reference flux $f_\mathrm{ref}^{(p)}$.
In this analysis, we set the reference flux to the observed pulsed flux:
\begin{equation}
f_{\mathrm{ref}}^{(p)}
=
\frac{1}{d_p}\sum_{c=1}^{n_c}\sum_{t=1}^{N}Y^{(c,p)}_t
-
\frac{1}{d_{p_{\mathrm{off}}}}\sum_{c=1}^{n_c}\sum_{t=1}^{N}Y^{(c,p_{\mathrm{off}})}_t,
\label{eq:f0_definition}
\end{equation}
where $p_{\mathrm{off}}=2$ denotes the off-pulse phase.
Therefore, the sky intensity and the point-source flux are calculated by
\begin{equation}
  (I^\ast,f^\ast)=\argmax_{I\ge0,\,f\ge0}\left[\log p(Y\mid I,f) - \mu_1 V_1(I) -\mu_2 V_2(I) - \mu_3 V_3(f) \right].
\label{eq:total_objective}
\end{equation}

\subsection{Optimization}\label{sec:optimization}

We maximize Eq.~(\ref{eq:total_objective}) using the limited-memory BFGS algorithm with bound constraints (L-BFGS-B), which requires the objective function and its gradient \citep{Byrd1995,Zhu1997}. We use the implementation provided by \texttt{scipy.optimize}
\citep{2020SciPy-NMeth} with bounds \(I_s\ge0\) and
\(f^{(q,p)}\ge0\).

The gradient of the log-likelihood term with respect to $I_s$ is
\begin{equation}
\frac{\partial \log p(Y\mid I,f)}{\partial I_s}
= \sum_{p=1}^{n_p}\sum_{c=1}^{n_c}\sum_{t=1}^{N}
m^{(c)}_{t,s}\left(\frac{Y^{(c,p)}_t}{E^{(c,p)}_t}-1\right)d_p .
\label{eq:grad_like_I}
\end{equation}
The gradient of the log-likelihood term with respect to $f^{(q,p)}$ is
\begin{equation}
\frac{\partial \log p(Y\mid I,f)}{\partial f^{(q,p)}}
= \sum_{c=1}^{n_c}\sum_{t=1}^{N}
\left(\frac{Y^{(c,p)}_t}{E^{(c,p)}_t}-1\right)
\left(\sum_{s=1}^{N} m^{(c)}_{t,s} I^{(q)}_{\mathrm{unit},s}\right)d_p.
\label{eq:grad_like_f}
\end{equation}
The gradient of the polarization-angle smoothness term is
\begin{equation}
\frac{\partial V_1(I)}{\partial I_{(i,j),\ell}}
=
\begin{cases}
2\left(2I_{(i,j),1}-I_{(i,j),2}-I_{(i,j),N_q}\right),
& \ell=1,\\
2\left(2I_{(i,j),\ell}-I_{(i,j),\ell+1}-I_{(i,j),\ell-1}\right),
& 2\le \ell\le N_q-1,\\
2\left(2I_{(i,j),N_q}-I_{(i,j),1}-I_{(i,j),N_q-1}\right),
& \ell=N_q.
\end{cases}
\label{eq:grad_V1_closed}
\end{equation}
Here the first and last cases mean the periodic boundary condition along the polarization-angle coordinate.

The gradient of the spatial smoothness term depends on the position of the spatial pixel because boundary pixels have fewer nearest neighbors.
We therefore write it separately for interior pixels, edge pixels excluding corners, and corner pixels.
For an interior pixel $(2\le i\le N_x-1,\,2\le j\le N_y-1)$,
\begin{equation}
\frac{\partial V_2(I)}{\partial I_{(i,j),\ell}}
=
2\Big(4I_{(i,j),\ell}-I_{(i+1,j),\ell}-I_{(i-1,j),\ell}
-I_{(i,j+1),\ell}-I_{(i,j-1),\ell}\Big).
\label{eq:grad_V2_interior}
\end{equation}
For edge pixels excluding corners,
\begin{align}
\frac{\partial V_2(I)}{\partial I_{(1,j),\ell}}
&=
2\Big(3I_{(1,j),\ell}-I_{(2,j),\ell}-I_{(1,j+1),\ell}-I_{(1,j-1),\ell}\Big),
&& (2\le j\le N_y-1), \label{eq:grad_V2_topedge}\\
\frac{\partial V_2(I)}{\partial I_{(N_x,j),\ell}}
&=
2\Big(3I_{(N_x,j),\ell}-I_{(N_x-1,j),\ell}
-I_{(N_x,j+1),\ell}-I_{(N_x,j-1),\ell}\Big),
&& (2\le j\le N_y-1), \label{eq:grad_V2_bottomedge}\\
\frac{\partial V_2(I)}{\partial I_{(i,1),\ell}}
&=
2\Big(3I_{(i,1),\ell}-I_{(i+1,1),\ell}-I_{(i-1,1),\ell}-I_{(i,2),\ell}\Big),
&& (2\le i\le N_x-1), \label{eq:grad_V2_leftedge}\\
\frac{\partial V_2(I)}{\partial I_{(i,N_y),\ell}}
&=
2\Big(3I_{(i,N_y),\ell}-I_{(i+1,N_y),\ell}
-I_{(i-1,N_y),\ell}-I_{(i,N_y-1),\ell}\Big),
&& (2\le i\le N_x-1). \label{eq:grad_V2_rightedge}
\end{align}
For corner pixels,
\begin{align}
\frac{\partial V_2(I)}{\partial I_{(1,1),\ell}}
&=
2\Big(2I_{(1,1),\ell}-I_{(2,1),\ell}-I_{(1,2),\ell}\Big),
\label{eq:grad_V2_corner11}\\
\frac{\partial V_2(I)}{\partial I_{(1,N_y),\ell}}
&=
2\Big(2I_{(1,N_y),\ell}-I_{(2,N_y),\ell}-I_{(1,N_y-1),\ell}\Big),
\label{eq:grad_V2_corner1N}\\
\frac{\partial V_2(I)}{\partial I_{(N_x,1),\ell}}
&=
2\Big(2I_{(N_x,1),\ell}-I_{(N_x-1,1),\ell}-I_{(N_x,2),\ell}\Big),
\label{eq:grad_V2_cornerM1}\\
\frac{\partial V_2(I)}{\partial I_{(N_x,N_y),\ell}}
&=
2\Big(2I_{(N_x,N_y),\ell}-I_{(N_x-1,N_y),\ell}
-I_{(N_x,N_y-1),\ell}\Big).
\label{eq:grad_V2_cornerMN}
\end{align}
The gradient of the point-source flux term is
\begin{equation}
\frac{\partial V_3(f)}{\partial f^{(q,p)}}
=
2\left(\sum_{q'=1}^{N_q} f^{(q',p)} - f_{\mathrm{ref}}^{(p)}\right).
\label{eq:grad_V3}
\end{equation}

%For each $k=1,\dots,K$, we define training--validation pair $k$ by using the $k$-th subset as the validation data and the remaining $K-1$ subsets as the training data.
%Thus, the procedure gives $K$ different training--validation pairs.
%We randomly divide detected events into $K$ subsets evenly to make $K$ training--validation data pairs.
%For each $k=1,\dots,K$, we form one training--validation pair by using the $k$-th subset as the validation data and the remaining $K-1$ subsets as the training data.
%For $k$-th pair $(k=1,\dots,K)$, the $k$-th subset is used as the validation data, and the remaining $K-1$ subsets are used as the training data.
%data where a pair consists of $k$-th subset and the remaining $K-1$ subsets respectively ($k=1,\dots,K$).
\subsection{Hyperparameter selection by cross-validation} \label{sec:hyperparameterselect}

We select the hyperparameter set ${\mu}=(\mu_1,\mu_2,\mu_3)$ using $K$-fold cross-validation.
For an IXPE event dataset, we randomly divide the detected events into $K$ subsets with approximately equal numbers of events to make the $K$ training--validation data pairs for cross-validation, where the $k$-th pair consists of the $k$-th subset as the validation data and the remaining $K-1$ subsets as the training data ($k=1,\dots,K$).
We use $K=5$ in this analysis.
We denote by $Y^{(c,p,k)}_{t}$ the number of events in detector voxel $t$ for DU $c$ and phase interval $p$ of the $k$-th validation data.

We evaluate each hyperparameter set ${\mu}$ in the three-dimensional hyperparameter space using the cross-validation score defined below, which is computed from the $K$ training--validation pairs.
For each hyperparameter set ${\mu}$ and pair $k$, we obtain the estimates $\hat I^{(k)}_s({\mu})$ and $\hat f^{(q,p,k)}({\mu})$ by deconvolving the $k$-th training data.
The expected detector counts are calculated from these estimates as
\begin{equation}
\hat E^{(c,p,k)}_t({\mu})
=
\left\{
\sum_{s=1}^{N}
m^{(c)}_{t,s}
\left[
\hat I^{(k)}_s({\mu})
+
\sum_{q=1}^{N_q}
\hat f^{(q,p,k)}({\mu}) I^{(q)}_{\mathrm{unit},s}
\right]
\right\}
d_p .
\label{eq:cv_expected_counts}
\end{equation}
We quantify the difference between $\hat E^{(c,p,k)}_t({\mu})$ and $Y^{(c,p,k)}_{t}$ by the root-mean-square error (RMSE) for the $k$-th pair:
\begin{equation}
\mathrm{RMSE}^{(k)}({\mu})
=
\sqrt{
\frac{1}{n_c n_p N}
\sum_{c=1}^{n_c}\sum_{p=1}^{n_p}
\sum_{t=1}^{N}
\left(\frac{1}{K-1}\,\hat E^{(c,p,k)}_{t}({\mu})-Y^{(c,p,k)}_{t}\right)^2
}.
\label{eq:rmse_fold}
\end{equation}
Here, the factor $1/(K-1)$ accounts for the sample size difference between training and validation data.
We calculate the cross-validation score by the mean over the $K$ pairs,
\begin{equation}
\overline{\mathrm{RMSE}}({\mu})
=
\frac{1}{K}\sum_{k=1}^{K}\mathrm{RMSE}^{(k)}({\mu}),
\label{eq:rmse_cv_mean}
\end{equation}
and also calculate the standard error of this mean by
\begin{equation}
\mathrm{SE}({\mu})
=
\sqrt{\frac{1}{K(K-1)}\sum_{k=1}^{K}\left(\mathrm{RMSE}^{(k)}({\mu})-\overline{\mathrm{RMSE}}({\mu})\right)^2}.
\label{eq:rmse_cv_sd}
\end{equation}

In cross-validation, hyperparameter sets with smaller values of the cross-validation score are generally preferred.
Choosing the set with the smallest score, however, can lead to weak regularization and increase the risk of overfitting to training data.
To reduce this risk, we adopt the one-standard-error (1-SE) heuristic
\citep{hastie2015statistical}.
The 1-SE heuristic defines a set of acceptable hyperparameter sets whose cross-validation scores are within one standard error of the minimum:
\begin{equation}
\overline{\mathrm{RMSE}}({\mu})
\le
\overline{\mathrm{RMSE}}({\mu}_{\min}) + \mathrm{SE}(\mu_{\min}),
\label{eq:one_se_inequality}
\end{equation}
where ${\mu}_{\min}$ denotes the hyperparameter set that gives the smallest value of $\overline{\mathrm{RMSE}}({\mu})$.
%We define the normalized deviation from the minimum cross-validation score as
By defining the normalized deviation from the minimum cross-validation score as
\begin{equation}
G({\mu})
=
\frac{
\overline{\mathrm{RMSE}}({\mu})
-
\overline{\mathrm{RMSE}}({\mu}_{\min})
}{
\mathrm{SE}(\mu_{\min})
},
\label{eq:G_def}
\end{equation}
the 1-SE acceptable region is expressed as $G({\mu})\le 1$.
Within this region, the 1-SE heuristic chooses a hyperparameter set of the strongest regularization: $G({\mu})=1$.
In the three-dimensional hyperparameter space, however, this condition does not specify a unique hyperparameter set.
To determine a single hyperparameter set, we therefore introduce an additional score to balance the contribution among the three hyperparameters ($\mu_1$, $\mu_2$, $\mu_3$):
\begin{equation}
S({\mu}) = w_1\log \mu_1 + w_2 \log \mu_2 + w_3 \log \mu_3,
\label{eq:regularization_strength}
\end{equation}
where the weights $w_i$ adjust the strength of log $\mu_i$ for $G(\mu)$.
We determine the weights as proportional to the change in $\log G({\mu})$ per unit change in $\log\mu_i$: $w_i = k \Delta \log G(\mu) / \Delta \log \mu_i \, (k:\mathrm{coefficient}; i = 1, 2, 3)$ because the relationship between $G(\mu)$ and $\mu_i$ is approximately linear on a log-log scale, as shown in Appendix~\ref{app:weight}.
With $w_1$ fixed at 1, $w_2$ and $w_3$ are determined as described in Appendix~\ref{app:weight}.

We use Bayesian optimization (BO) to reduce the computational cost of searching the three-dimensional hyperparameter space \citep{BO_Shahriari}.
Since the 1-SE acceptable region is defined relative to
$\overline{\mathrm{RMSE}}({\mu}_{\min})$, the BO search is performed in two steps.
In the first step, we minimize $\overline{\mathrm{RMSE}}({\mu})$ to estimate ${\mu}_{\min}$.
In the second step, we minimize the objective
\begin{equation}
g({\mu}) = \left(G({\mu})-1\right)^2 - \eta S({\mu}).
\label{eq:bo_objective_impl}
\end{equation}
The first term penalizes deviations from $G({\mu})=1$, corresponding to the boundary of the 1-SE acceptable region.
The second term helps to select a single hyperparameter set near this boundary while balancing the contributions of the three regularization parameters.
We choose $\eta>0$ sufficiently small that the second term does not dominate the first term, and set $\eta=0.1$ in this analysis (see Appendix~\ref{app:eta_displacement} for the setting of $\eta$).
We define the selected hyperparameter set as ${\mu}^{\ast}=\arg\min_{{\mu}} g({\mu})$.

\subsection{Response matrix} \label{subsec:response}
The coupled spatial and polarization-angle response matrix $m^{(c)}_{t,s}$ is not available in a ready-to-use form in the standard IXPE calibration products.
We therefore make the response matrix from simulations.
The $s$-th column of $m^{(c)}$ represents the detector-space count distribution produced by emission with unit intensity from a single sky voxel $s$.
Thus, the full response matrix can be calculated by separately simulating the event distribution in detector space for each sky voxel and stacking these distributions as the columns of the matrix.
We generate these event distributions using the IXPE simulation and analysis framework \texttt{ixpeobssim} \citep{Baldini2022}.

We assume that the point-source responses at different positions within the ROI can be represented by a single response template.
This approximation also neglects spatial variations of the exposure and vignetting across the ROI.
We therefore simulate only one template response and approximate the responses
for other sky voxels by shifting this template in the two spatial directions
and by circularly shifting it in the polarization-angle direction.
This approximation is expected to be adequate for the small field of view used
here ($3.40^\prime\times3.40^\prime$), because off-axis PSF variations are
modest and the mirror aberration is reported to be much smaller than the PSF
half-power diameter across the field of view \citep{ixpeobssim_irf_doc}.

We generate the template response by simulating a point source with 100\% linear polarization and an input polarization angle of $\psi=0$.
For simplicity, we adopt a monochromatic photon energy of 3~keV and use the \texttt{ixpeobssim} calibration response set specified by \texttt{irfname=ixpe:obssim20240701:v13}.
Polarization leakage is not included in the calibration response set used here and is therefore omitted from our response matrix.
The source and pointing directions are both set to the Crab pulsar coordinates, placing the simulated source at the image center.
We generate $2\times10^{7}$ simulated events for each DU so that Monte Carlo uncertainty in the resulting template is negligibly small compared with the statistical uncertainty in the observed data.
The simulated events are binned with the same spatial pixel size and polarization-angle binning as the analysis detector cube, but on a larger $69\times69$ spatial grid to provide a margin for shifting the template before extracting the central $35\times35$ spatial grid.

For each sky voxel $s=(u,\ell)$, we calculate the corresponding column of the response matrix from the template response.
Let $u_0=(i_0,j_0)$ denote the spatial pixel of the injected point source in the template, and let $\ell_0$ denote the polarization-angle bin corresponding to the input polarization angle $\psi=0$.
For $u=(i,j)$, the template is shifted by $(i-i_0,j-j_0)$ pixels in the two spatial directions and circularly shifted by $\ell-\ell_0$ bins in the polarization-angle direction.
After shifting the template, we extract the central $35\times35\times16$ region to match the analysis detector cube.
The cropped detector distribution is assigned to the $s$-th column of $m^{(c)}$.
After shifting and cropping, we normalize each column $s$ so that the total counts summed over all detector voxels of the three DUs equal unity:
\begin{equation}
\sum_{c=1}^{n_c}\sum_{t=1}^{N} m^{(c)}_{t,s}=1.
\end{equation}
With this normalization, $m^{(c)}_{t,s}$ can be interpreted as the probability that a photon originating from sky voxel $s$ is recorded in detector voxel $t$ of DU $c$.
%\hl{TBD}

The resulting response matrix has dimensions $N\times N$ for each DU, where $N=35\times35\times16$ in this work.
When stored in dense format using double-precision floating-point numbers, each response matrix requires approximately 2.86~GiB for each DU, or approximately 8.6~GiB for all three DUs.
The storage size of the matrix can be reduced by representing it in sparse format because many elements are numerically negligible.
For a given sky voxel $s$, the value of $m^{(c)}_{t,s}$ is very small for detector voxels $t$ far from the sky position of $s$.
To reduce the computational cost, we therefore set elements smaller than $10^{-5}$ to zero and then renormalize each sky-voxel column to preserve the normalization introduced above.
The size of the response matrix is reduced to approximately 0.5--0.6~GiB per DU, or approximately 1.6~GiB for all three DUs.
In the numerical implementation, the response matrices are stored on the GPU in a sparse format and reused for repeated deconvolution runs during hyperparameter selection.

\section{Results}
\subsection{Selected hyperparameters} \label{subsec:crossvalid}

\begin{figure*}[!htbp]
\centering
\includegraphics[width=10cm]{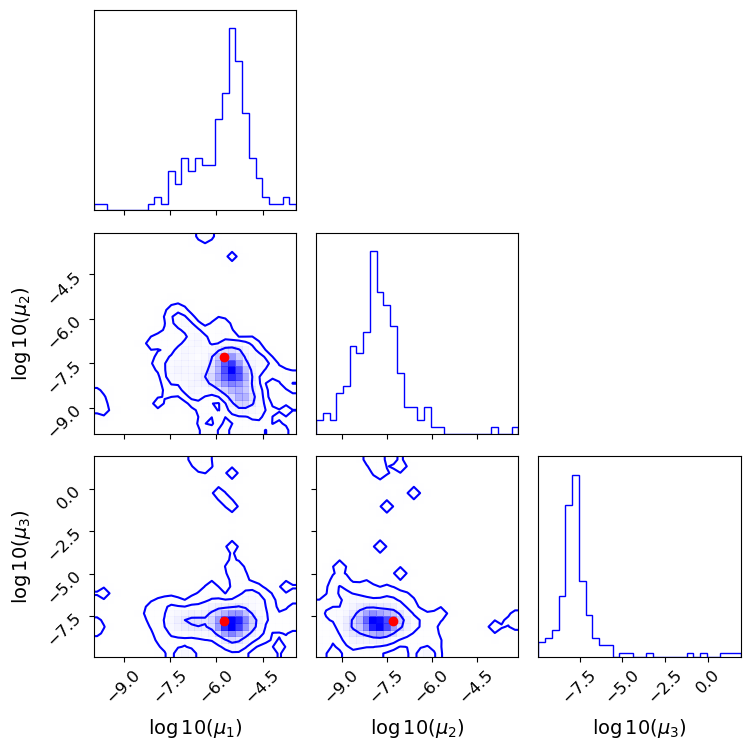}
\caption{Corner plot of the hyperparameter sets evaluated in the second BO run, shown in $(\log_{10}\mu_1,\log_{10}\mu_2,\log_{10}\mu_3)$ space.
The diagonal panels show one-dimensional histograms of the hyperparameter values sampled in the 170 subsequent optimization trials.
The off-diagonal panels show Gaussian-smoothed histograms, with contours enclosing 50\%, 80\%, and 95\% of the smoothed histogram mass.
The first 30 randomly sampled startup trials are excluded from the plot for clarity.
The red marker denotes the trial that minimizes the objective $g({\mu})$ (Eq.~\ref{eq:bo_objective_impl}).}
\label{fig:crossval_1sigma}
\end{figure*}

We select the hyperparameter set ${\mu}=(\mu_1,\mu_2,\mu_3)$ using the cross-validation procedure described in Section~\ref{sec:hyperparameterselect}. The weights $(w_2,w_3)$ in Eq.~(\ref{eq:regularization_strength}) are estimated as described in Appendix~\ref{app:weight}; we obtain $w_2=0.86$ and $w_3=3.45$.
The BO search is implemented with \texttt{Optuna} \citep{Akiba2019}.
We search $\mu_1 \in [10^{-10},10^{-2}]$, $\mu_2 \in [10^{-10},10^{-2}]$, and $\mu_3 \in [10^{-10},10^{2}]$, with each hyperparameter sampled on a logarithmic scale.
In the first BO run, we minimize $\overline{\mathrm{RMSE}}({\mu})$ over 100 trials, consisting of 30 randomly sampled startup trials and 70 subsequent trials.
This yields $\overline{\mathrm{RMSE}}({\mu}_{\min})=3.02$ at
${\mu}_{\min}=(7.55\times10^{-9},2.98\times10^{-10},4.56\times10^{-9})$.
In the second BO run, we minimize the objective function $g({\mu})$ defined in Eq.~(\ref{eq:bo_objective_impl}), using the value of ${\mu}_{\min}$ obtained from the first run.
We perform 200 trials, consisting of 30 randomly sampled startup trials and 170 subsequent trials.
Figure~\ref{fig:crossval_1sigma} shows the hyperparameter sets evaluated in the second BO run, and the red marker indicates the selected hyperparameter set ${\mu}^{\ast}=({1.82\times10^{-6}},{5.04\times10^{-8}},{1.68\times10^{-8}})$.

\subsection{Deconvolved image and polarization maps} \label{subsec:results}
\begin{figure*}[!htbp]
\centering
\includegraphics[width=\textwidth]{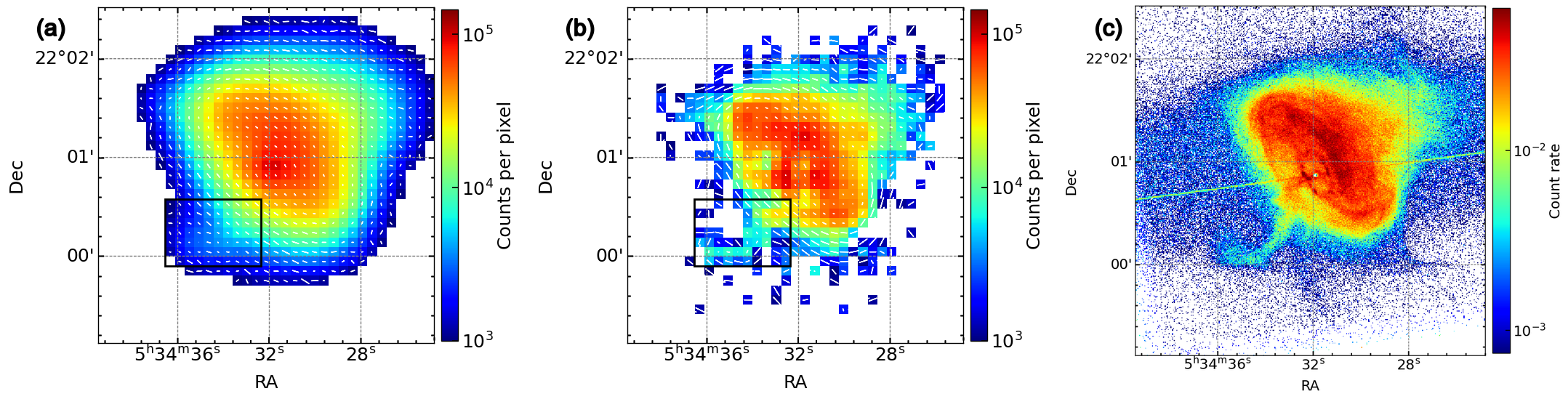}
\caption{Comparison of the Crab Nebula images in the 2--8~keV band.
(a) Intensity map overlaid with magnetic-field orientations obtained from
the basic binned polarization analysis using the \texttt{ixpeobssim}
\texttt{pcube} products for the IXPE observation (ObsID 01001099).
(b) Map of the reconstructed nebular component $I_s$ obtained from our deconvolution, summed over the polarization-angle coordinate.
In panels (a) and (b), colors indicate the total intensity per spatial pixel,
and white vectors indicate the magnetic-field orientations, with lengths
proportional to the polarization degree.
Pixels with fewer than 1000 counts are masked in the IXPE polarization maps.
The black rectangle marks the jet region $\mathcal{A}$ analyzed in Fig.~\ref{fig:jet_polar_combine}.
(c) Chandra ACIS-S image of the Crab Nebula (ObsID 23539) in the same
2--8~keV band and sky region, shown for comparison of the spatial morphology.
}
\label{fig:main_maps}
\end{figure*}

Using the hyperparameter set ${\mu}^\ast$ determined in Section~\ref{subsec:crossvalid}, we reconstruct the polarization map of the Crab Nebula.
For comparison, we also perform a basic polarization analysis using  \texttt{pcube} for the same event selection and ROI.
Figure~\ref{fig:main_maps} (a,b) compares the intensity and polarization maps obtained from the basic analysis and from our deconvolution.
For the basic analysis, we make the map from the full event set, including all pulse phases, and combine the three DUs into a single map.
For our deconvolution result, we show the reconstructed nebular component $I_s$, summed over the polarization-angle coordinate.
This component corresponds to the phase-independent nebular emission in our model and therefore does not include the phase-dependent point-source component.
%In both panels, the colors indicate the intensity per spatial pixel and the overlaid vectors indicate the magnetic field orientations on the sky, which is perpendicular to the polarization angle.
In both panels, the colors indicate the intensity per spatial pixel and the overlaid vectors show the magnetic field orientation at each pixel (Appendix~\ref{app:polarization}), defined as perpendicular to the corresponding polarization angle.
For both panels, pixels with intensities below 1000 counts are masked for clarity.
The deconvolved map shows sharper spatial structure than the basic binned map, particularly around the jet.
Figure~\ref{fig:main_maps} (c) shows a 2--8~keV Chandra ACIS-S image (ObsID 23539) of the same sky region for comparison of the spatial morphology.

\begin{figure*}[!htbp]
\centering
\includegraphics[width=0.75\textwidth]{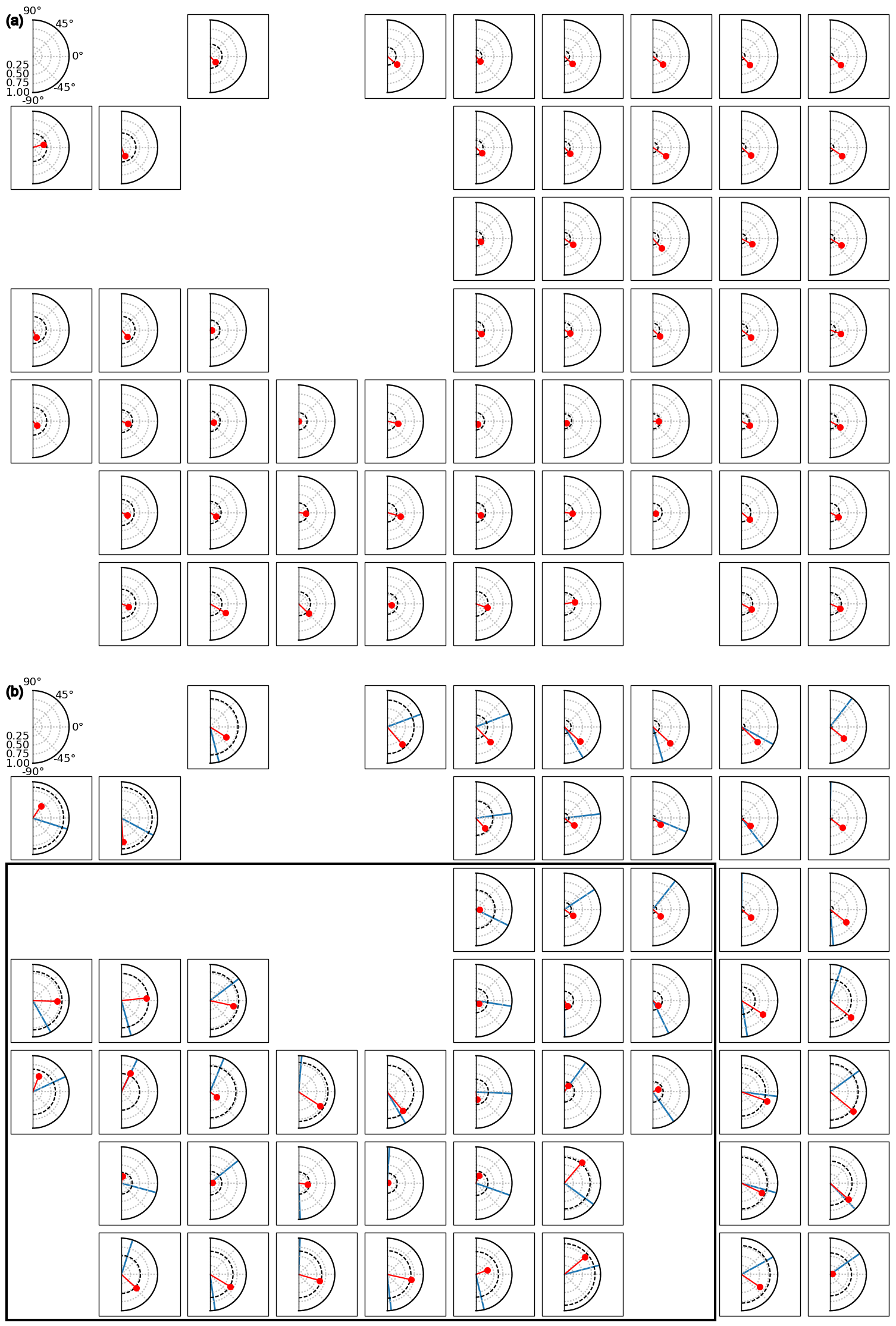}
\caption{(a) Pixel-wise polarization results from the basic analysis.
Each panel corresponds to one spatial pixel within the jet region $\mathcal{A}$ shown in Fig.~\ref{fig:main_maps}.
In each panel, the red vector indicates the magnetic field orientation, with its length proportional to the polarization degree.
The dashed semicircle indicates the 90\% MDP.
The angular and radial scales are common to all panels and are shown in the top-left reference plot.
For clarity, tick labels are shown only in the reference plot.
(b) Pixel-wise polarization results obtained from our deconvolution.
Each panel corresponds to one spatial pixel within the jet region $\mathcal{A}$ shown in Fig.~\ref{fig:main_maps}.
In each panel, the red vector indicates the magnetic field orientation, with its length proportional to the reconstructed polarization degree (Appendix~\ref{app:polarization}).
The blue line shows the intensity-gradient orientation derived from the deconvolved intensity map; this quantity is defined in Section~\ref{sec:stat_eval}.
The dashed semicircle indicates the empirical 90\% MDP estimated from unpolarized Monte Carlo simulations through the deconvolution procedure (Section~\ref{sec:stat_eval}).
The black rectangle marks the jet body--to--tip subregion $\mathcal{S}$ used for the statistical analysis in Section~\ref{sec:stat_eval}.
}
\label{fig:jet_polar_combine}
\end{figure*}

We next examine the polarization pattern in the jet region $\mathcal{A}$ indicated by the black rectangle in Fig.~\ref{fig:main_maps}.
%Figure~\ref{fig:jet_polar_pcube} shows the pixel-wise polarization results from the standard analysis in the jet region $\mathcal{A}$ shown in Figure~\ref{fig:main_maps}, where the polarization degree, the inferred magnetic-field orientation, and the 90\% minimum detectable polarization (MDP) are shown.
Figure~\ref{fig:jet_polar_combine} (a) shows the pixel-wise polarization results from the basic analysis in the jet region $\mathcal{A}$ denoted in Fig.~\ref{fig:main_maps}, including the polarization degree, magnetic-field orientation, and 90\% minimum detectable polarization (MDP).
The red vector indicates the magnetic field orientation, which is perpendicular to the polarization angle.
The dashed semicircle indicates the 90\% MDP obtained from the \texttt{ixpeobssim} \texttt{pcube} analysis, where it is calculated using the standard count-based method \citep{KISLAT201545}.

%\begin{figure*}[!htbp]
%\centering
%\includegraphics[width=\textwidth]{pcube_pol_jet_v2.png}
%\caption{Pixel-wise polarization results from the standard analysis.
%Each panel corresponds to one spatial pixel within the jet region $\mathcal{A}$ shown in Fig.~\ref{fig:main_maps}.
%In each panel, the red vector indicates the magnetic field orientation, with its length proportional to the polarization degree.
%The dashed semicircle indicates the 90\% MDP.
%The angular and radial scales are common to all panels and are shown in the top-left reference plot.
%For clarity, tick labels are shown only in the reference plot.
%}
%\label{fig:jet_polar_pcube}
%\end{figure*}

Figure~\ref{fig:jet_polar_combine} (b) shows the pixel-wise polarization results obtained from our deconvolution in the jet region shown in Fig.~\ref{fig:main_maps}.
Compared with the map from the basic analysis, the change in the magnetic field orientation from the jet body to the tip is more apparent in the deconvolved map.
Because the deconvolved intensity values are inferred through the reconstruction and are not direct event counts, the standard count-based MDP is not directly applicable.
Instead, the dashed semicircle in Fig.~\ref{fig:jet_polar_combine} (b) shows the empirical pixel-wise 90\% MDP estimated from unpolarized Monte Carlo simulations described in Section~\ref{sec:stat_eval}.
The black rectangle marks the jet body--to--tip subregion, denoted by $\mathcal{S}$, which is used for the statistical analysis in Section~\ref{sec:stat_eval}.
The blue lines in Fig.~\ref{fig:jet_polar_combine} (b) indicate the intensity-gradient orientation derived from the deconvolved intensity map.

The magnetic field orientation appears approximately perpendicular to the jet-flow direction in the jet body (upper right) and more nearly parallel to the jet near the tip (lower left).
Within the subregion $\mathcal{S}$, the magnetic field orientation appears to be approximately perpendicular to the local intensity-gradient orientation. We statistically quantify this relation in Section~\ref{sec:stat_eval}.
The change in magnetic field orientation relative to the jet-flow direction is broadly consistent with previous spatially resolved analyses of the Crab Nebula, which reported a transition of the magnetic field orientation from an approximately perpendicular orientation at the jet body area to roughly parallel along the jet \citep{Wong_2024}.
It needs to be noted that their analysis accounts for instrumental effects including the PSF structure, polarization leakage, and the energy dependence of the modulation factor.
Our response model uses a simplified PSF and omits the other two effects (Section~\ref{subsec:response}), which may cause systematic errors in our reconstructed polarization maps.

\subsection{Statistical evaluation of the relation between magnetic-field and intensity-gradient orientations in the subregion $\mathcal{S}$} \label{sec:stat_eval}
The empirical pixel-wise 90\% MDP values shown in Fig.~\ref{fig:jet_polar_combine} (b) are estimated from unpolarized Monte Carlo simulations generated under the null hypothesis of zero polarization.
To generate the null distribution, we make an unpolarized sky intensity from the reconstructed nebula sky intensity $I_{(u,\ell)}$.
At each spatial pixel $u$, we preserve the total intensity summed over the polarization-angle coordinate and redistribute it uniformly over the $N_q$ polarization-angle bins.
Specifically, we define
\begin{equation}
I_{\mathrm{null},(u,\ell)}
=
\frac{1}{N_q}I_{\mathrm{tot}}(u),
\qquad \ell=1,\dots,N_q,
\label{eq:unpol_cube_def}
\end{equation}
where
\begin{equation}
I_{\mathrm{tot}}(u) = \sum_{\ell=1}^{N_q} I_{(u,\ell)}.
\end{equation}
We compute the expected detector counts for each DU from $I_{\mathrm{null}}$ using the response matrix,
\begin{equation}
E^{(c)}_{\mathrm{null},t}
=
\sum_s
m_{t,s}^{(c)}
I_{\mathrm{null},s}.
\end{equation}
Here, we ignore the pulsar component because the contribution of the pulsar to the subregion $\mathcal{S}$ is small compared with the nebula component.
We generate $N_{\mathrm{MC}}=5000$ Monte Carlo realizations of the detector counts by Poisson sampling with means $E^{(c)}_{\mathrm{null},t}$.
We index these realizations by $r=1,\dots,N_{\mathrm{MC}}$ and denote the corresponding detector counts by $Y^{(c,r)}_{\mathrm{null},t}$ for DU $c=1,2,3$:
\begin{equation}
Y^{(c,r)}_{\mathrm{null},t}
\sim
\mathrm{Poisson}\!\left(E^{(c)}_{\mathrm{null},t}\right).
\label{eq:null_poisson_realization}
\end{equation}
For each realization $r$, we reconstruct the nebular sky intensity $I$ from the corresponding detector counts for the three DUs using our deconvolution procedure.
Here, we use the polarization-angle and spatial smoothness priors with $\mu_1^\ast$ and $\mu_2^\ast$, respectively, which are obtained from the cross-validation of the original observed data.
The prior of flux constraint is omitted.
We define the 90\% MDP at pixel $u$, $\mathrm{MDP}_{90}(u)$, as the 90th percentile of the reconstructed polarization degrees obtained at that pixel from the $N_{\mathrm{MC}}$ unpolarized realizations.

In the jet body--to--tip subregion $\mathcal{S}$, the reconstructed magnetic field orientation appears to be approximately perpendicular to the intensity-gradient orientation (Section~\ref{subsec:results}).
Since the photon counts in individual pixels remain limited, we need to evaluate the significance of this relation statistically over the pixels in $\mathcal{S}$.
We therefore define a concentration statistic for the angle differences between the magnetic field orientation and the intensity-gradient orientation in $\mathcal{S}$, and evaluate its significance by comparing the observed statistic with those of empirical null distributions obtained under the two null hypotheses described below.

We define the intensity-gradient orientation from the deconvolved intensity map.
For each spatial pixel $u=(i,j)$, we estimate the two gradient components using the Prewitt operators $\mathcal{P}_x$ and $\mathcal{P}_y$ along the two spatial directions \citep{Prewitt1970}.
We then define the gradient-direction angle as
\begin{equation}
\phi_{\mathrm{grad}}(u) =
\arg\left[
\mathcal{P}_xI_{\mathrm{tot}}(u)
+i \mathcal{P}_yI_{\mathrm{tot}}(u)
\right].
\label{eq:grad_angle_def}
\end{equation}
We define the intensity-gradient orientation
$\theta_{\mathrm{grad}}(u)\in\left[-\frac{\pi}{2},\frac{\pi}{2}\right)$
by treating opposite gradient directions as equivalent:
\begin{equation}
\theta_{\mathrm{grad}}(u) =
\left[\phi_{\mathrm{grad}}(u)+\frac{\pi}{2}\right]\bmod \pi - \frac{\pi}{2}.
%\in\left[-\frac{\pi}{2},\frac{\pi}{2}\right)
\label{eq:axial_map}
\end{equation}
Let $\theta_{\mathrm{B}}(u)\in\left[-\frac{\pi}{2},\frac{\pi}{2}\right)$ denote the magnetic field orientation, which is perpendicular to the polarization angle.
For each pixel $u$ in the jet body--to--tip subregion $\mathcal{S}$, we define the signed angle difference between the intensity-gradient orientation and the magnetic field orientation as
\begin{equation}
\Delta(u) =
\left[\theta_{\mathrm{grad}}(u)-\theta_{\mathrm{B}}(u)+\frac{\pi}{2}\right]\bmod \pi - \frac{\pi}{2},
\label{eq:delta_axial_def}
\end{equation}
where $\Delta(u)\in[-\pi/2,\pi/2)$.

We use the mean resultant length, which is a standard measure of concentration in circular statistics \citep{MardiaJupp2000}, as a statistic for the concentration of the angle differences $\Delta(u)$.
Because the directional information is not equally reliable in all pixels, we introduce weights $w(u)$ designed to emphasize pixels with stronger intensity gradients, higher polarization degrees, and larger event counts:
\begin{equation}
w(u) \propto \left\|\nabla I(u)\right\|\, \mathrm{PD}(u)\, N(u),
\label{eq:w_def}
\end{equation}
where
\begin{equation}
\left\|\nabla I(u)\right\|
=
\sqrt{\left[\mathcal{P}_xI_{\mathrm{tot}}(u)\right]^2+\left[\mathcal{P}_yI_{\mathrm{tot}}(u)\right]^2},
\label{eq:grad_norm_def}
\end{equation}
and $\mathrm{PD}(u)$ and $N(u)$ are the polarization degree and the number of events at pixel $u$, respectively.
This gives the weighted mean resultant length
\begin{equation}
\bar{R}_{w} =
\left|
\frac{\sum_{u\in\mathcal{S}} w(u)\,\exp\!\bigl[2i\Delta(u)\bigr]}
{\sum_{u\in\mathcal{S}} w(u)}
\right|.
\label{eq:wmrl_def}
\end{equation}
Here, the angle differences are doubled so that they can be treated as 2$\pi$-periodic angles in calculating the mean resultant length.

We compare the observed concentration statistic with the same statistic of Monte Carlo simulations under two null hypotheses.
%We evaluate the observed value of this concentration statistic using Monte Carlo simulations under two null hypotheses.
The first is the unpolarized null hypothesis, by which we test whether a concentration as large as the observed one can arise from Poisson fluctuations in the case of no polarization.
%This hypothesis tests whether a concentration as large as the observed one can arise from Poisson fluctuations and the reconstruction procedure in the absence of intrinsic polarization.
The second is a polarized null hypothesis in which the polarization angle is randomized independently at each spatial pixel while preserving the total intensity and polarization degree.
This hypothesis tests whether the observed concentration can arise when the polarization orientation has no relation to the local intensity-gradient orientation.
For both null hypotheses, we generate detector counts using the response matrix and Poisson sampling, and apply the same deconvolution procedure as described above for the MDP calculation.
%The Monte Carlo simulations and reconstruction procedure are the same as those described above for the MDP calculation.

%For each Monte Carlo realization $r$ and spatial pixel $u$, we randomly select an integer shift $k_u^{(r)}$ from $0,\ldots,N_q-1$, with all shifts equally likely, and define
%\begin{equation}
%I_{\mathrm{rand},(u,\ell)}^{(r)}
%=
%I_{(u,\ell')},
%\end{equation}
%where $\ell'=(\ell+k_u^{(r)})\bmod N_q$.
%For each randomized sky realization, we compute the expected detector counts for each DU using the same response matrix,
%\begin{equation}
%E_{\mathrm{rand},t}^{(c,r)}
%=
%\sum_s
%m_{t,s}^{(c)}
%I_{\mathrm{rand},s}^{(r)},
%\end{equation}
%and generate the corresponding detector counts as
%\begin{equation}
%Y_{\mathrm{rand},t}^{(c,r)}
%\sim
%\mathrm{Poisson}
%\left(E_{\mathrm{rand},t}^{(c,r)}\right).
%\label{eq:random_pa_poisson}
%\end{equation}
%Each realization is then deconvolved using the same procedure and the fixed hyperparameters $\mu_1^\ast$ and $\mu_2^\ast$ used for the zero-polarization Monte Carlo simulations.

\begin{figure*}[!htbp]
\centering

\begin{minipage}{0.75\textwidth}
    \centering
    \includegraphics[width=0.8\textwidth]{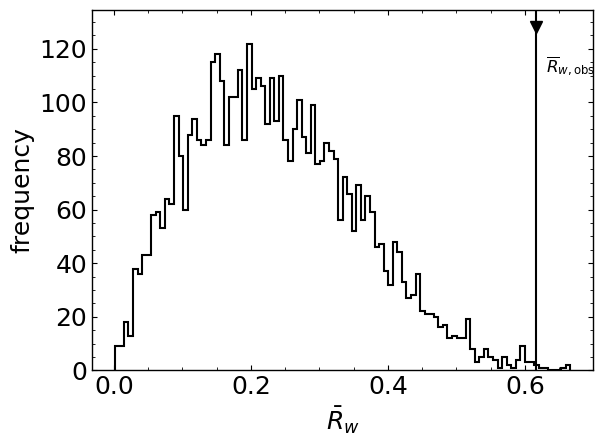}\\
    \textbf{(a)}
\end{minipage}

\vspace{0.5cm}

\begin{minipage}{0.75\textwidth}
    \centering
    \includegraphics[width=0.8\textwidth]{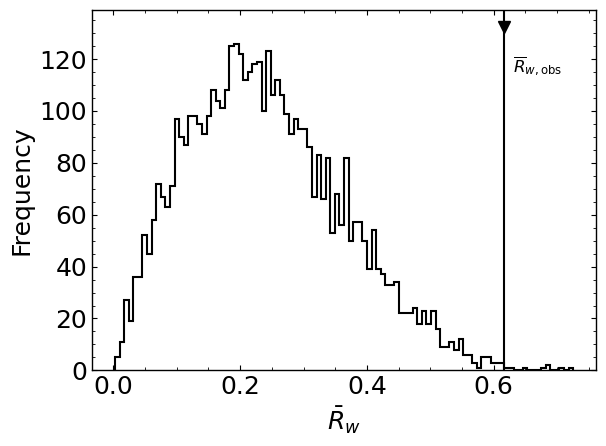}\\
    \textbf{(b)}
\end{minipage}

\caption{
Distributions of the weighted mean resultant length $\bar{R}_w$ for the
jet body--to--tip subregion $\mathcal{S}$ under two null hypotheses.
(a) Distribution obtained from $N_{\mathrm{MC}}=5000$ realizations under the
unpolarized null hypothesis.
(b) Distribution obtained from $N_{\mathrm{MC}}=5000$ polarized realizations
in which the polarization angle is randomized independently at each spatial
pixel while preserving the total intensity and polarization degree.
In both panels, the vertical line indicates the observed value,
$\bar{R}_{w,\mathrm{obs}}$, measured from the data.
}
\label{fig:MRL}
\end{figure*}

Figure~\ref{fig:MRL} shows the distributions of $\bar{R}_w$ obtained from
$N_{\mathrm{MC}}=5000$ realizations under the two null hypotheses,
together with the observed value $\bar{R}_{w,\mathrm{obs}}$.
Panel (a) shows the case of unpolarized null hypothesis, and panel (b) shows that of the polarized null hypothesis with randomized polarization angles.
We define the one-sided $p$-value as
\begin{equation}
p = \frac{N_{+}}{N_{\mathrm{MC}}},
\label{eq:pvalue_def}
\end{equation}
where $N_{+}$ is the number of null realizations satisfying
$\bar{R}_{w,\mathrm{null}} \ge \bar{R}_{w,\mathrm{obs}}=0.61$.
For the unpolarized null hypothesis, we obtain $N_{+}=6$ and $p=0.0012$.
For the polarized null hypothesis with randomized polarization angles,
we obtain $N_{+}=9$ and $p=0.0018$.

%To apply standard circular statistics, we double this angle difference($\in \left[-{\pi},{\pi}\right)$):
%\begin{equation}
%\varphi(u) = 2\Delta(u) .
%\label{eq:phi_double_def}
%\end{equation}
%We then evaluate a one-sided Monte Carlo tail probability,
%\begin{equation}
%p_{\mathrm{null}} = \frac{1}{N_{\mathrm{MC}}}\sum_{n=1}^{N_{\mathrm{MC}}}
%\mathbb{I}\!\left(\rho^{(n)}_{\mathrm{null}} \le \rho_{\mathrm{obs}}\right),
%\label{eq:mc_pvalue}
%\end{equation}
%(TBD: report $\rho_{\mathrm{obs}}$ and $p_{\mathrm{null}}$), where $\mathbb{I}(\cdot)$ is the indicator function.
%In bright regions with high photon statistics, it also yields lower MDP values than the standard \texttt{pcube}-based analysis.

%The magnetic field orientation from an approximately perpendicular orientation at the jet body area to roughly parallel along the jet.

\section{Discussion and Conclusion} \label{sec:discuss}

We applied a MAP-based polarization-resolved deconvolution to the first IXPE observation of the Crab Nebula.
The deconvolution yields polarization-resolved maps with sharper spatial structure than the basic (\texttt{pcube}-based) analysis, making the jet morphology more clearly identifiable.
In the jet body--to--tip subregion $\mathcal{S}$, the reconstructed magnetic field orientation shows a statistically significant relation to the local intensity-gradient orientation.
The weighted mean resultant length of the angular differences ($\bar{R}_w$) is larger than those expected under the null hypotheses with no polarization and randomized polarization angles ($p=0.0012$ and $0.0018$, respectively), indicating that the recovered polarization pattern at jet regions is unlikely to be explained by statistical fluctuations.

A previous spatially resolved IXPE analysis by \citet{Wong_2024} reported that the magnetic-field orientation relative to the jet axis changes from approximately perpendicular in the jet body to approximately parallel along the jet.
In our analysis, the magnetic field orientation varies with the local intensity-gradient orientation across region $\mathcal{S}$, maintaining an approximately perpendicular relation.
As discussed by \citet{Wong_2024}, the observed change in magnetic field orientation along the jet may be related either to kink instabilities along the jet, which can strongly deflect a magnetized flow \citep{Mignone2013}, or to collision with a dense medium near the jet tip, which may deflect the jet and reorganize the local magnetic field geometry.
A comparison between the polarization pattern recovered in the deconvolved jet map and polarization patterns predicted by theoretical models will be an interesting direction for future study.

We use $N_q=16$ equal-width polarization-angle bins for all spatial pixels.
We chose this binning empirically by considering the smoothness of the reconstructed image and polarization-angle distributions, the computational cost, and the angular sampling needed to recover the polarization signal.
Coarser binning does not adequately sample the polarization-angle distribution, whereas finer binning increases the computational cost and makes the deconvolution less stable against Poisson fluctuations, producing spurious modulation and increasing the reconstructed polarization degree.
The same effect in unpolarized Monte Carlo realizations also increases the empirical MDP.
In the outer regions of the nebula (Fig.~\ref{fig:main_maps}), including the jet region (Fig.~\ref{fig:jet_polar_combine}), both the reconstructed polarization degree and the empirical MDP tend to be larger than those obtained from the basic analysis.
Varying the polarization-angle binning according to the local photon counts may improve reconstruction stability.

The response model used in our analysis involves several approximations.
We approximate the position-dependent PSF across the ROI by spatially shifting a single template and omit polarization leakage. Spatial variations of exposure and vignetting are also neglected.
These simplifications may affect the reconstructed spatial polarization pattern.
We also adopt a monochromatic response at 3~keV, using the modulation factor at this energy throughout the nebula.
Because the Crab Nebula's spectrum varies spatially and the modulation factor depends on energy, the effective modulation factor averaged over the detected photon spectrum can differ between sky positions.
Our monochromatic approximation does not account for these differences and may therefore cause spatially varying systematic errors in the reconstructed polarization degree.
%However, these approximations are not required by our deconvolution framework.
Since our method is formulated in polarization-angle count space, the Stokes-based polarization-leakage treatment adopted by \citet{Wong_2024}, based on the model of \citet{Dinsmore_2024}, cannot be applied.

The polarization leakage can instead be incorporated into the simulated detector response used to construct the response matrix.
The monochromatic approximation can also be avoided by making the response matrix assuming a spectrum at each sky position.
These improvements can be made without changing our deconvolution framework and are important tasks for future work.
Since multiple IXPE observations of the Crab are available, merging these datasets and applying the deconvolution to the combined data may improve significance in the jet and enable a more detailed evaluation of the polarization structure we find.

These results show that polarization-resolved deconvolution with smoothness priors can enhance the scientific interpretability of IXPE data by improving polarization maps and revealing spatial structures that are difficult to isolate in the basic analysis.

\begin{acknowledgments}
This work was supported by JSPS KAKENHI Grant Number JP19H05609.
We acknowledge the use of ChatGPT, DeepL, and Google Translate for translation assistance, language editing and grammar checking.
\end{acknowledgments}

%\begin{contribution}
%%This section gives authors the space to recognize author contributions. The text inside this environment is NOT counted towards the total word quanta. At a minimum, manuscripts a    re expected to include this text:

%All authors contributed equally to the Terra Mater collaboration.

%\end{contribution}

\facilities{IXPE}

\software{ixpeobssim \citep{Baldini2022},
          NumPy \citep{harris2020array},
          SciPy \citep{2020SciPy-NMeth},
          Astropy \citep{astropy:2022},
          PyTorch \citep{NEURIPS2019_bdbca288},
          Optuna \citep{Akiba2019}}


%and the local linear fits in log--log space used to determine $\alpha_i$.
%We determine the weights as proportional to the change in $\log G({\mu})$ per unit change in $\log\mu_i$: $w_i = k \Delta \log G(\mu) / \Delta \log \mu_i \, (i = 1, 2, 3)$.

%%: for each hyperparameter, we hold the other two fixed, define the normalized deviation from the corresponding minimum, and average it over the grid values of the other two hyperparameters to
\appendix
\section{Weights in the score function $S(\mu)$}\label{app:weight}

To estimate $\Delta\log G(\mu)/\Delta\log\mu_1$, we first define a normalized deviation as a function of $\mu_1$ while holding $\mu_2$ and $\mu_3$ fixed:
\begin{equation}
{G}_1(\mu_1;  \mu_2,\mu_3)
=
\frac{
\overline{\mathrm{RMSE}}(\mu_1,\mu_2,\mu_3)
-
\overline{\mathrm{RMSE}}(\mu_1^\dagger,\mu_2,\mu_3)
}{
\mathrm{SE}(\mu_1^\dagger,\mu_2,\mu_3)
},
\label{eq:G_tilde_def}
\end{equation}
where $\mu_1^\dagger$ is defined as
\begin{equation}
\mu_1^\dagger
=
\argmin_{\mu_1}
\overline{\mathrm{RMSE}}(\mu_1,\mu_2,\mu_3).
\label{eq:mu_star_def}
\end{equation}
We calculate $G_1$ at ($\mu_1$, $\mu_2$, $\mu_3$) on a three-dimensional logarithmically spaced grid with $N_i=10$ points in each axis $(i=1,2,3)$, with $\mu_1\in[10^{-8.5},10^{-4}]$,
$\mu_2\in[10^{-9.5},10^{-5}]$, and
$\mu_3\in[10^{-8.5},10^{-7}]$.
We average $G_1$ over all grid values of $\mu_2$ and $\mu_3$:
\begin{equation}
\overline{G}_1(\mu_1)
=
\frac{1}{N_2N_3}
\sum_{a=1}^{N_2}
\sum_{b=1}^{N_3}
{G}_1
\left(
\mu_1;
\mu_2^{(a)},\mu_3^{(b)}
\right),
\label{eq:G1_average}
\end{equation}
where $\mu_2^{(a)}$ denotes the $a$-th grid value of $\mu_2$ $(a=1,\ldots,N_2)$, and $\mu_3^{(b)}$ denotes the $b$-th grid value of $\mu_3$ $(b=1,\ldots,N_3)$.
We apply the same procedure to $\mu_2$ and $\mu_3$ and obtain $\overline{G}_2(\mu_2)$ and $\overline{G}_3(\mu_3)$, respectively.

\begin{figure*}[!htbp]
\centering
\includegraphics[width=0.5\columnwidth]{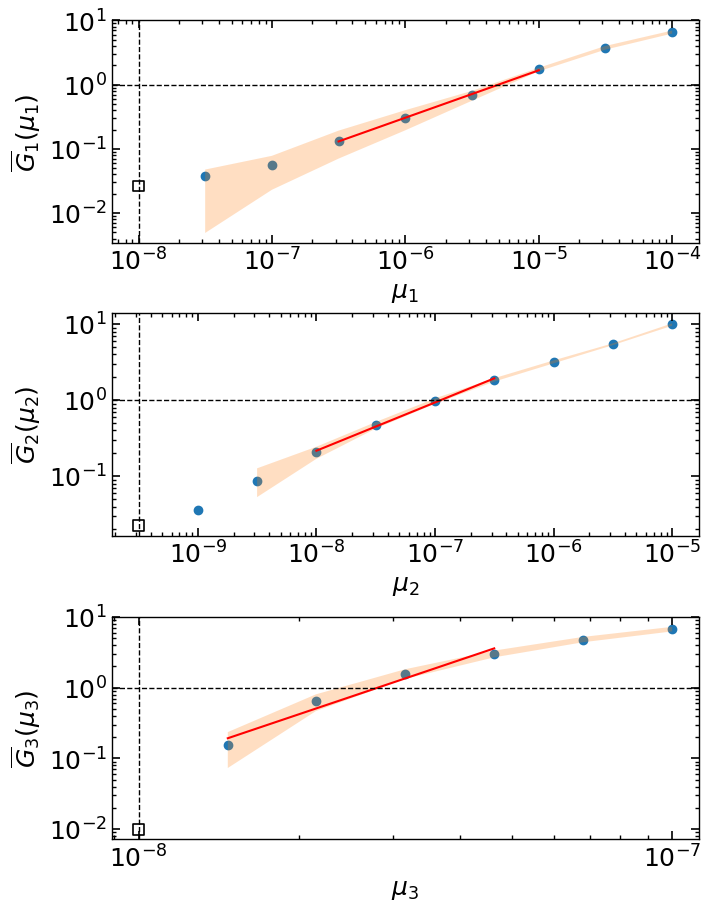}\\
\caption{
$\overline{G}_i(\mu_i)$ used to determine the hyperparameter weights $(w_2,w_3)$.
Blue points show $\overline{G}_i(\mu_i)$ and shaded bands indicate the interquartile ranges (25--75\%) of ${G}_i$.
The red lines show the best-fit lines with the function~(\ref{eq:Gtilde_powerlaw_fit}) using the four grid points around $\overline{G}_i=1$, indicated by the horizontal dashed line.
The open square in each panel marks the minimum of $\overline{G}_i(\mu_i)$, and the vertical dashed line indicates the corresponding value of $\mu_i$.
Points with $\mu_i$ smaller than this value are omitted for clarity.
}
\label{fig:Gtilde_marginals}
\end{figure*}

Figure~\ref{fig:Gtilde_marginals} shows $\overline{G}_i(\mu_i)$ for $i=1,2,3$.
The relationships between $\overline{G}_i(\mu_i)$ and $\mu_i$ are approximately linear on a log--log scale.
We therefore fit each relationship with
\begin{equation}
\log \overline{G}_i(\mu_i)
\approx
\alpha_i\log\mu_i+b_i,
\label{eq:Gtilde_powerlaw_fit}
\end{equation}
where $\alpha_i$ is the slope in the $\log\mu_i$--$\log\overline{G}_i$ plane and provides an estimate of $\Delta\log G(\mu)/\Delta\log\mu_i$ ($i=1,2,3$).
We fix $w_1=1$ and define the remaining weights as
\begin{equation}
w_2=\alpha_2/\alpha_1,\qquad
w_3=\alpha_3/\alpha_1.
\label{eq:weights_def}
\end{equation}
For each hyperparameter, we fit the function~(\ref{eq:Gtilde_powerlaw_fit}) to the four grid points around $\overline{G}_i=1$.
The red lines in Fig.~\ref{fig:Gtilde_marginals} show the best-fit results.
The fitted slopes are
$\alpha_1=0.738$, $\alpha_2=0.634$, and $\alpha_3=2.545$, which give
$w_2=0.86$ and $w_3=3.45$.

\section{Setting of $\eta$}
\label{app:eta_displacement}
The minimum point of $g({\mu}) = \left[G({\mu})-1\right]^2 -\eta S({\mu})$, $\mu^*$, is different from that of $\left[G({\mu})-1\right]^2$, $\mu^0$, corresponding to 1-SE boundary ($G=1$).
We choose $\eta$ so that this difference in the selected hyperparameters can remain within a numerical precision $\delta$.
For simplicity, we first vary $\mu_1$ while holding $\mu_2$ and $\mu_3$ fixed.
The stationary condition for $g$ with respect to $\mu_1$ is
\begin{equation}
\frac{\partial g}{\partial\mu_1}
=
2(G-1)\frac{\partial G}{\partial\mu_1}
-\eta\frac{w_1}{\mu_1} = \frac{1}{\mu_1}\left[ 2(G-1) \frac{\partial G}{\partial x_1} -\eta w_1 \right]
=0,
\label{eq:eta_stationary_mu1}
\end{equation}
where $x_1=\ln\mu_1$.
Near the 1-SE boundary, we approximate $G$ to the first order in $x_1$:
\begin{equation}
G(x_1)
\simeq
1+\left.
\frac{\partial G}{\partial x_1}
\right|_{x_1=x^0_{1}}
(x_1-x^0_{1}).
\label{eq:eta_local_G}
\end{equation}
Approximating $\frac{\partial G}{\partial x_1}|_{x_1=x^0_{1}} \simeq \frac{\partial G}{\partial x_1}|_{x_1=x^*_{1}} = a_1,$
we obtain
$x^*_1-x^0_1
\simeq
\eta w_1/{2a_1^2}
$, where $x^*_1$ is the minimum point of $g$.
Requiring this difference to be smaller than $\delta$ gives $\eta < 2a_1^2\delta/w_1$.
We approximate $a_1$ by the slope $\frac{\partial \bar{G_1}}{\partial \log \mu_1}=\bar{G_1} \frac{\partial \log \bar{G_1}}{\partial \log \mu_1} = \alpha_1$ at $\bar{G_1}=1$ obtained in Appendix~\ref{app:weight}.
$\mu_2$ and $\mu_3$ are estimated similarly.
We set $\delta=0.2\ln 10$, which corresponds to roughly one grid in Fig.~\ref{fig:crossval_1sigma}.
This gives upper bounds $0.50,0.43,1.73$ for $i=1,2,3$, respectively.
We therefore set $\eta=0.1$.

%< \delta$, where $\delta$ is the maximum allowed shift in $x_1$.
%For each hyperparameter,
%\begin{equation}
%x^*_{i}-x^0_{i}
%\simeq
%\frac{\eta w_i}{2a_i^2},
%\qquad
%x_i=\ln\mu_i,
%\label{eq:eta_displacement_general}
%\end{equation}
%where $x^0_{i}=\ln\mu^0_{i}$ corresponds to $G=1$ and $x^*_{i}=\ln\mu^*_{i}$ corresponds to a minimum of $g$ along the $\mu_i$ direction.
\section{Polarization degree and angle of reconstructed image}
\label{app:polarization}
For each spatial pixel $u$, we calculate the polarization degree and polarization angle from the reconstructed intensity distribution along the polarization-angle coordinate as follows.
Let $\psi_\ell$ denote the center of the $\ell$-th polarization-angle bin
and let $\hat{I}_{u,\ell}$ denote the reconstructed nebular intensity in
that bin.
We define
$I(u) = \sum_{\ell=1}^{N_q} \hat{I}_{u,\ell}, Q(u) = \sum_{\ell=1}^{N_q} \hat{I}_{u,\ell}
        \cos 2\psi_\ell$, and  $U(u) = \sum_{\ell=1}^{N_q} \hat{I}_{u,\ell}
        \sin 2\psi_\ell .$
The polarization degree and polarization angle are then calculated as
\begin{align}
{\rm PD}(u)
&=
\frac{\sqrt{Q(u)^2+U(u)^2}}{I(u)}, \\
\psi(u)
&=
\frac{1}{2}\arg\!\left[Q(u)+iU(u)\right].
\end{align}
Here, the angle is calculated using the \texttt{arctan2} function.

\clearpage
\bibliography{ixpedeconvref_apj_clean}{}
\bibliographystyle{aasjournalv7-fixed}



\end{document}